# Climate modification directed by control theory

Wang Liang

(Department of Control Science and Control Engineering, Huazhong University of Science and Technology, WuHan, 430074, P.R.China, E-mail: wl@smail.hust.edu.cn)

**[Abstract]** Climate modification measures to counteract global warming receive some more new attentions in these years. Most current researches only discuss the impact of these measures to climate, but how to design such a climate regulator is still unknown. This paper shows the control theory could give the systematic direction for climate modification. But the control analyzing also reveals that climate modifications should only be regarded as a last-ditch measure.

## 1 Introduction

IPCC (Intergovernmental panel on climate change) announce its Fourth Assessment Report in 2007. Adaptation and mitigation options are suggested to avoid all climate change impacts. Besides the mitigation method like afforestation in this report (1), some more ambitious measures are also reconsidered in these years, including putting up space shields that cover billions of square meters, using chemicals to reflect sunlight or increase Earth's cloud cover, stimulating massive growth of phytoplankton in the oceans (2-7). Assessment of climate mitigation or manipulation has also been reviewed under various framings including economics, risk, politics, and environmental ethics (8).

But all these work still can't answer these questions: How to adjust the measures according to prediction or actual impact of active manipulation? Can the project be readily reversed if it goes awry? How to evaluate the effect of one measure in conjunction with other methods? Obviously, a systematic 'strategy' to modify the climate system is still absent. Here we will shows the control theory could meet this requirement.

## 2 Basic analyses for climate control design

Control theory is an interdisciplinary branch of engineering and mathematics, which deals with influencing the behavior of dynamical systems. The controller is designed to manipulate the inputs to a system to obtain the desired effect on the output of the system.

Selecting the proper model is the first step for control research. There are three kinds of climate models, simple climate model, earth models of intermediate complexity (EMICs) and global climate models (GCMs) (9). Their complexities increase in turn. Many research shows the results given by most complex GCMS can also be captured by much simple model. So we use simple climate model to design the control law and then verify this law in sophisticated GCM models.

Here we select a simple climate model considering the atmosphere and the underlying surface (10,11,12):

$$\begin{cases} C_A \dfrac{dT_A}{dt} = (1-\alpha_A)a[1+(1-a)\alpha_S]Q + \varepsilon\delta T_A^4 - 2\varepsilon T_s^4 + H(T_s - T_A) \\ C_S \dfrac{dT_s}{dt} = (1-\alpha_A)(1-a)(1-\alpha_s)Q - \delta T_S^4 + \dfrac{4}{3}\varepsilon\delta T_A^4 - H(T_S - T_A) \end{cases} \quad (1)$$

$C_A = 4.6 \times 10^7 Wm^{-2}K^{-1}$ $\quad C_S = 2.97 \times 10^8 Wm^{-2}K^{-1}$ $\quad Q = 342 Wm^{-2}$,

$\delta = 5.67 \times 10^{-8} Wm^{-2}K^{-4}$, $a = 0.241, \varepsilon = 0.812, \alpha_s = 0.132, \alpha_A = 0.250$,

$H = 5.944 Wm^{-2}K^{-1}$.

Here $T_A, T_S$ is average air temperature and surface air temperature. The description of other parameters could be found in (10). This model is zero-dimensional, representing global mean and vertically integrated conditions. In the model, energy is exchanged between the surface, atmosphere and space by short and long-wave radiation and by latent and sensible heat.

Then we should determine the appropriate control input or said control measure. The overwhelming majority of climate manipulation proposals aim to alter radiative energy fluxes, either by increasing the amount of outgoing infrared radiation through reduction of atmospheric CO2, or by decreasing the amount of absorbed solar radiation through an increase in albedo. Here we will consider this measure as control input. In principle, the use of space-based solar shields has significant advantages over other options (Fig.1). Because solar shields effect a 'clean' alteration of the solar constant, their side effects would be both less significant and more predictable than for other albedo modification schemes (13).

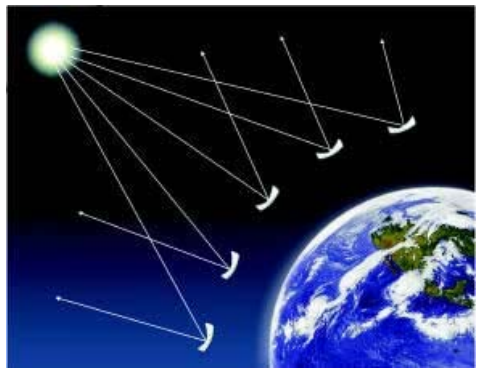

Fig.1. Space-based albedo modification (a picture from Internet)

Many current climate control proposals belong to open loop, which only consider the impact of measures. But how to regulate these measures according to the modification effect is not mentioned, which may result in the system unstable. For example, Aerosols produced in the lower stratosphere can counteract some of the effects of global warming. But excessive aerosols may overly offset the warming and bring adverse impacts. To avoid the problems of the open-loop, most current controller uses feedback to control states or outputs of a dynamical system. The name 'close-loop feedback' comes from the information path in the system: process inputs have an effect on the process outputs, which is measured with sensors and processed by the controller; the result (the control signal) is used as input to the process, closing the loop. For aerosols measures, the amount of aerosols injected in the air should be adjusted according to its impact to climate.

Here $U$ denotes the control input. As a close-loop control system, it should be the function of $T_A, T_S$. So mark it as $U(T_A, T_S)$. As an albedo modification control, it could be added in formula (1) as follows:

$$\begin{cases} C_A \dfrac{dT_A}{dt} = (1-\alpha_A)a[1+(1-a)\alpha_S]Q[1-U(T_A,T_S)] + \varepsilon\delta T_A^4 - 2\varepsilon T_s^4 + H(T_s - T_A) \\ C_S \dfrac{dT_s}{dt} = (1-\alpha_A)(1-a)(1-\alpha_s)Q[1-U(T_A,T_S)] - \delta T_S^4 + \dfrac{4}{3}\varepsilon\delta T_A^4 - H(T_S - T_A) \end{cases} \quad (2)$$

Here $U$ is the percent of energy fluxes that the control implementer could curtain off. A particular issue is the requirement for a control system to perform properly in the presence of input and state constraints. In the physical world every signal is limited. For example, the energy we can manipulate is very limited comparing with climate system. So the designed controller should avoid sending control signals that cannot be followed by the physical system. Here set the control constraint as:

$$U \in [0, 3\%] \quad (3)$$

Then we need decide a main control aim. For climate system, most related researchers think that manipulations need not be aimed at changing the environment, but rather may aim to maintain a desired environmental state against perturbations—either natural or anthropogenic. This aim could be regarded as 'state regulator' problem and 'constant tracing' problem in control theory. The following parts will give the detailed discussion for these two control problems.

## 3 Climate state regulator design

The state regulator is an operator driving the system to equilibrium state by minimum cost. The cost function could be the energy consumption, total time, etc. We could define the climate regulator problem as follows.

For given system (1), its equilibrium state is:

$T_A = 270.2K$, $T_S = 288.0K$

Then for some 'temporary' perturbations, the temperature raise to:

$T_A = 274K$, $T_S = 292K$

The system structure has no changes. We want to drive the climate system to original state as soon as possible. This problem could be defined as:

Initial state: $[T_A(t_0), T_S(t_0)] = (270.2, 288.0)$

Final state: $[T_A(t_f), T_S(t_f)] = (274, 292)$

Cost function: $J = \int_{t_0}^{t_f} 1 dt$

Control aims: $J = \min(J)$

Many different design methods are available this problem. The risks posed by climate control are sufficiently novel that, in general, the relevant biological and geophysical science is too uncertain to allow quantitative.

In control engineering, typically a simpler mathematical model is chosen in order to simplify calculations; otherwise the true system dynamics can be so complicated that a complete model is impossible, especially for climate system. But no real physical system truly behaves like the series of differential equations used to represent it mathematically. So the climate control system must always have some robustness property. A robust controller is such that its properties do not change much if applied to a system slightly different from the mathematical one used for its synthesis. Now there are several robust design methods like $H_\infty$ design.

Here we select the cell mapping design method, which is simple, robust and easy to be understood (14-16). This method belongs to Bellman dynamic programming method. We can depict this method in an easy way: it divides the state space into small rectangle regions called cells. The control input is also discretized. Then a search algorithm is applied to assign only control input for each rectangle region/cell to meet the optimal requirement.

The main operation of cell mapping design method is described as follows:

1) Select interesting state region. $T_A \in [268, 276]$, $T_S \in [286, 294]$.

2) Construct cell. We divide the $T_A$ and $T_S$ into $2^6 = 64$ equal pieces, so there are $64 \times 64 = 4096$ cells.

3) Determine the control input set U. The input set is divide into 8 equal pieces: $U = \{0, 0.37\%, 0.75\%, 1.13\%, 1.5\%, 1.87\%, 2.25\%, 2.62\%, 3\%\}$.

4) Obtain Mappings from every cell with the entire set of control information U. For each cell, we need calculate $N_U = 9$ number of cell transitions. Here select one quarter as the integral time. The related cost is also obtained in this step.

5) Decide the only control information U for each cell. Dynamic programming is employed to identify this information. The search algorithm associates each cell with a control action that maps the cell to a cell trajectory with the optimal cost. Discrete Optimal Control Table (DOC) is obtained to represent the discrete global optimal control solution. All the optimum trajectories from every possible initial condition in the cell state space can be generated once this database is built. The detailed algorithm could be found in Hsu's original paper. In our instance, there are 3271 controllable cells. A cell is said to be controllable if there exists a sequence of controls which could bring this cell to the target cell. The controllable region and DOC table for this example is shown in Fig.2:

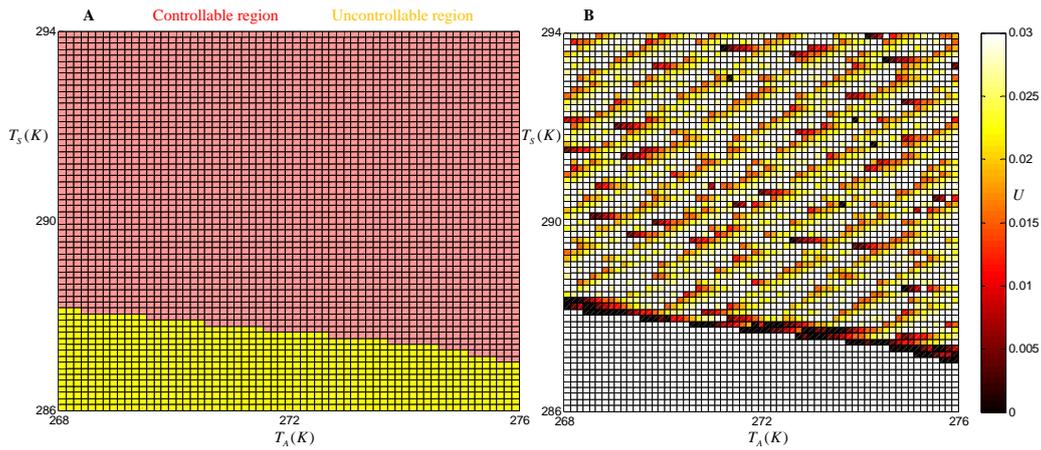

Fig.2. (A) Controllable region, (B) Optimal control table, the control value for each cell/region is represented by different color.

6) The optimal sequence of control is readily obtainable from DOC. Because DOC stores the singe control input for every cell, in operation process, the system only need read the input value from DOC according to the feedback state. The cell optimal solution of our system is shown in fig.3:

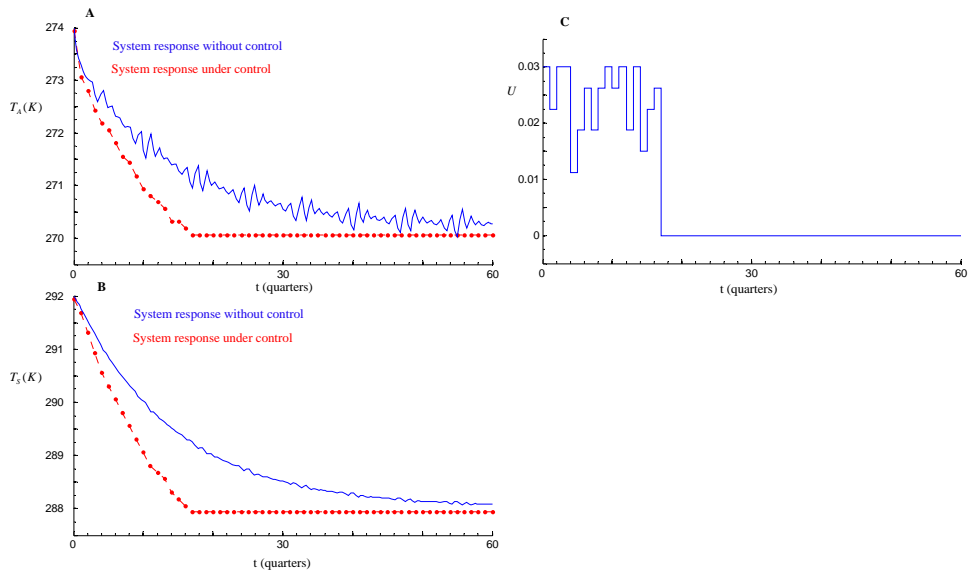

Fig.3. (A) , (B) Response of system , (C) Control input

In Fig.3.A, B, we could find the system need about 60 quarters to resume to equilibrium state without any control. But if we add the control, it only needs about 20 quarters to resume.

## 4 Global warming offset design

Now the main topic for climate modification is the compensation of $CO_2$ emission. It's a 'tracing problem'.

Here assume the $CO_2$ reach 560ppm in 2150, two times of concentration before industrialization. Then $CO_2$ will not increase anymore. We simulate this scene in a complex

GCM model GISS Ⅱ (17,18). The global air surface temperature will increase from ~286.88K to a new equilibrium ~290.3K after 2090, which is shown in Fig.4.

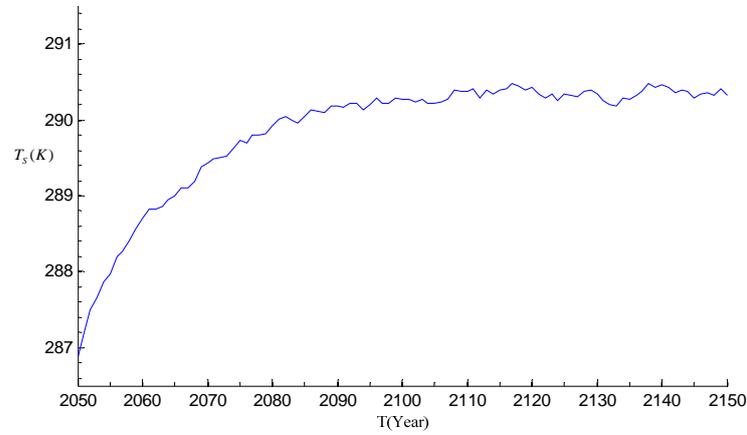

Fig.4. Global temperature from 2050 to 2150

The temperature in 2065 is ~288.2K, which is similar to current condition. So we use it as the reference scene. The temperature changes between 2056 and 2100 are shown in Fig.5.

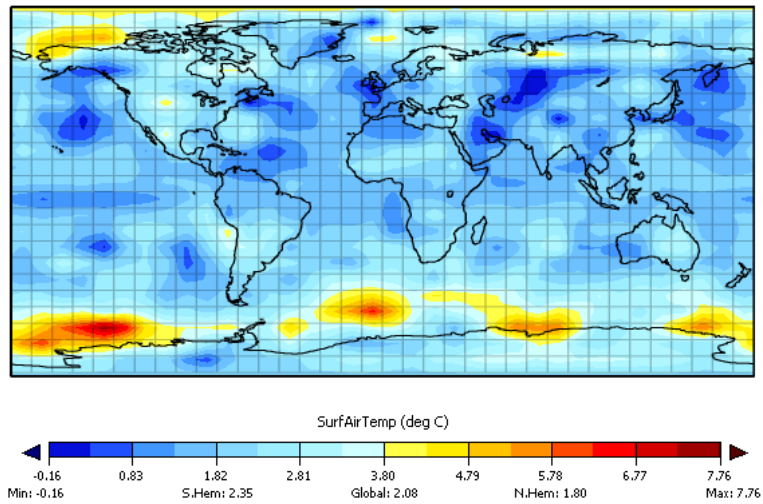

Fig.5. Temperature changes between 2056 and 2100

Here the control aim could be defined as reducing the temperature from equilibrium state ~290.3k to the state of 2065 (~288.2K) using least time.

We also design the control law by simple climate model (1), where the impact of CO2 is featured by $\varepsilon$. We could adjust this parameter to simulate this climate system. If $\varepsilon = 0.8408$, the new equilibrium state is: $T_A = 271.56K$, $T_S = 290.34K$, this temperature is close to the equilibrium state of GCM models. Because there is no a corresponding air temperature in GCM model, we simply assume $T_A = T_s - 18.5$ in GCM. We begin the control in 2100, its $T_S = 290.3K, T_A = 271.8K$. This problem could be defined as:

Initial state: $[T_A(t_0), T_S(t_0)] = (271.8, 290.3)$

Final state: $[T_A(t_f), T_S(t_f)] = (269.7, 288.2)$

Cost function: $J = \int_{t_0}^{t_f} 1 dt$

Control aims: $J = \min(J)$

We still use the cell mapping method to design this control law. Here select the interesting area: $T_A \in [269, 273]$, $T_S \in [287, 291]$, integral time:1 year, other setting is similar to the operation in paragraph 3. The DOC table of this system is shown in Fig.6:

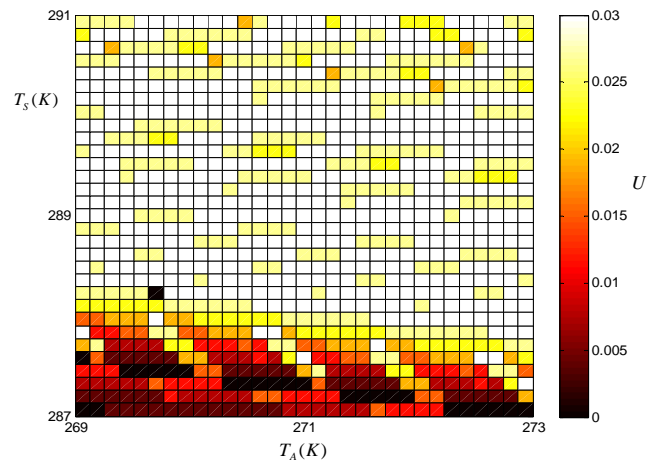

Fig.6. Optimal control table for global warming offset

Then we verify this control law in GCM model. In every year, we choose the control input from DOC table according to the average temperature of former year. The global average temperature reaches the control aim in 2105, which is shown in Fig.7:

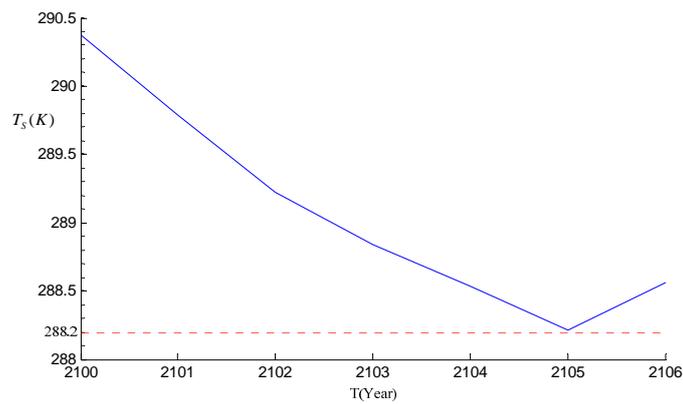

Fig.7. Global temperature from 2050 to 2150

If there is no a control strategy, we may not offset the warming so precisely. The control will end in 2106. Unlike the regulator problem, the temperature will rise again in 2106. So the control should continue according to the temperature of 2106. The DOC table should also be adjusted if CO2 concentration markedly changes. If concentration gets back to a normal level, the control strategy could be converted to state regulator control.

The temperature changes between reference year 2065 and 2105 are shown in Fig.8:

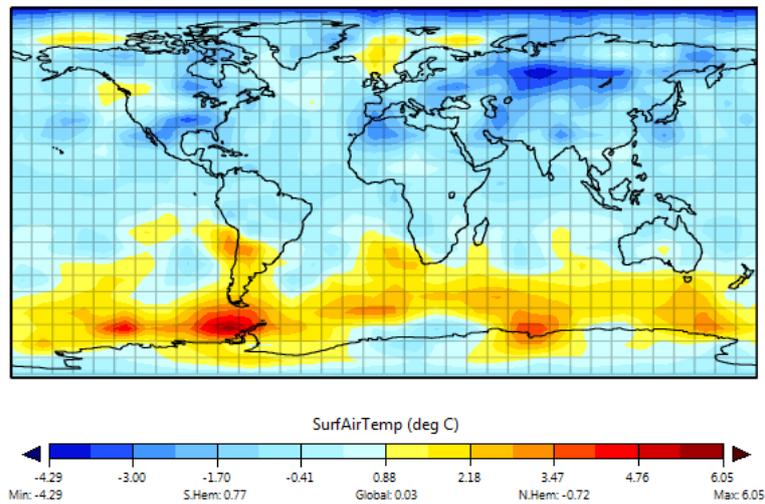

Fig.8. Temperature changes between 2056 and 2105

From Fig.8, we could find the compensation is not very uniform, though the difference of averages temperatures between 2056 and 2105 is only 0.03K. To avoid this problem, multiple compensation measures or more precise simple climate model and more complex design method could be considered in the future.

## 5 Conclusions

The control theory could make the climate modification more reliable and safe. But real climate system is a complex nonlinear, decentralized, strong coupling, distributed-parameters system, which still can't be well grasped by current control theory. The climate modification provides a new big challenge for control research. Moreover, the efficiency of control design strongly relies on how well we know the controlled system. But our knowledge about climate is still very limited. So we'd better test this technology in re-building the environment of Mars, but not on the Earth. Now the more reasonable and direct method may be still to curtail emissions of greenhouse gases.